\documentclass[prl,twocolumn,amsmath,amssymb,superscriptaddress,prl,floatfix]{revtex4}
\usepackage{color} 
\usepackage{graphicx}
\usepackage{dcolumn}
\usepackage{bm}
\usepackage{longtable}
\usepackage{textcomp}
\usepackage{epstopdf}
\usepackage{makeidx}     

\def\text{\mathrm}

\begin{document}

\setlength{\parskip}{0 pt}

\bibliographystyle{unsrt}

\title{X-Ray Detection of Transient Magnetic Moments Induced by a Spin Current in Cu}

\author{R. Kukreja}
\email{roopali.kukreja@gmail.com}
\affiliation{SLAC National Accelerator Laboratory, 2575 Sand Hill Road, Menlo Park, CA 94025, USA}
\affiliation{Department of Materials Science and Engineering, Stanford University, CA 94305, USA}

\author{S. Bonetti}
\affiliation{SLAC National Accelerator Laboratory, 2575 Sand Hill Road, Menlo Park, CA 94025, USA}
\affiliation{Department of Physics, Stanford University, CA 94305, USA}

\author{Z. Chen}
\affiliation{SLAC National Accelerator Laboratory, 2575 Sand Hill Road, Menlo Park, CA 94025, USA}
\affiliation{Department of Physics, Stanford University, CA 94305, USA}

\author{D. Backes}
\affiliation{Physics Department, New York University, 4 Washington Place, New York, NY 10003, USA}

\author{Y. Acremann}
\affiliation{Laboratorium f\"ur Festk\"orperphysik, ETH Z\"urich, HPF C 5, Otto-Stern-Weg 1, Z\"urich 8093, Switzerland }

\author{J. Katine}
\affiliation{HGST, a Western Digital Company, 3403 Yerba Buena Road, San Jose, CA 95135, USA}

\author{A.D. Kent}
\affiliation{Physics Department, New York University, 4 Washington Place, New York, NY 10003, USA}

\author{H. A. D\"urr}
\affiliation{SLAC National Accelerator Laboratory, 2575 Sand Hill Road, Menlo Park, CA 94025, USA}

\author{H. Ohldag}
\affiliation{SLAC National Accelerator Laboratory, 2575 Sand Hill Road, Menlo Park, CA 94025, USA}

\author{J. St\"ohr}
\email{stohr@slac.stanford.edu}
\affiliation{SLAC National Accelerator Laboratory, 2575 Sand Hill Road, Menlo Park, CA 94025, USA}

\begin{abstract}
We have used a MHz lock-in x-ray spectro-microscopy technique to directly detect changes of magnetic moments in Cu due to spin injection from an adjacent Co layer. The elemental and chemical specificity of x-rays allows us to distinguish two spin current induced effects. We detect the creation of transient magnetic moments of $3\times 10^{-5}$\,$\mu_\mathrm{B}$ on Cu atoms within the bulk of the 28\,nm thick Cu film due to spin-accumulation. The moment value is compared to predictions by Mott's two current model. We also observe that the hybridization induced existing magnetic moments on Cu interface atoms are transiently increased by about 10\% or $4\times 10^{-3}$\,$\mu_\mathrm{B}$. This reveals the dominance of spin-torque alignment over Joule heat induced disorder of the interfacial Cu moments during current flow.
\end{abstract}

\maketitle

One of the new paradigms in magnetism research is the use of spin currents to read and write static magnetic bits via the giant magneto-resistance (GMR) \cite{Baibach1988} and spin transfer torque effects \cite{Slonczewski1996, stohr-siegmann}. Spin currents are also believed to play a key role in the ultrafast manipulation of the magnetization by femtosecond optical pulses, like in all optical switching \cite{Stanciu2007,Radu2011,Graves:13}. They exist even during current flow through non-magnetic materials consisting of atoms with large spin-orbit coupling such as Pt, leading to spin accumulation through the spin Hall or Rashba effects \cite{ohno:2005,Miron2010}. The presence of spin currents is typically revealed through current or voltage dependent measurements, but an atomic level understanding of the detailed spin dependent scattering mechanisms requires techniques that can directly probe the electronic structure at the nanoscale.

In this letter we report the direct detection of the transient magnetization in a non magnet (Cu) caused by injection of spin polarized current from an adjacent ferromagnet (Co). This is accomplished by a significant advancement in the sensitivity of scanning transmission microscope (STXM) which is achieved by time dependent modulation of a spin current synchronized with x-ray pulses to produce a sensitivity corresponding to the magnetic moment of about 50 Fe atoms. Using this technique, we were able to measure an extremely small transient x-ray magnetic circular dichroism (XMCD) effect in the non-magnetic Cu layer. 

\begin{figure} [!t]
\includegraphics[width=0.9\columnwidth]{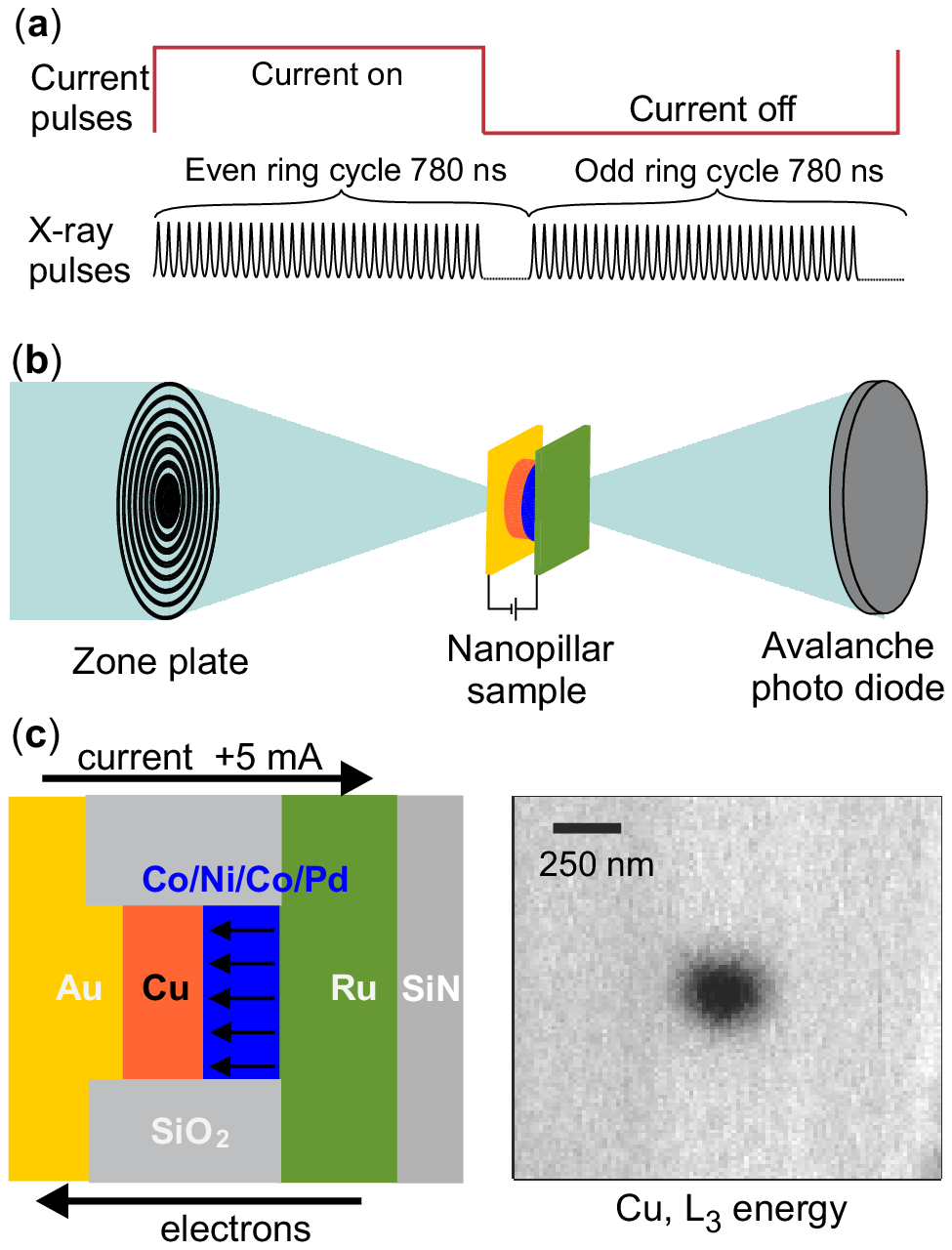}
\caption{ (a) Timing of current pulses, x-ray pulses and data collection periods, as discussed in the text. (b) Schematic of the x-ray microscopy measurements. The x-ray spot size at the sample was 35\,nm and the transmitted x-rays were detected by an avalanche photo diode. Images were recorded by raster scanning of the sample. (c) The sample consisted of a nanopillar of 240\,nm diameter containing a ferromagnetic multilayer with perpendicular magnetization direction, as discussed in the text. Current to the pillar was supplied by Au and Ru contact leads, as shown. The current is defined as positive when flowing from Cu to the ferromagnet, corresponding to electron flow in the opposite direction. At the bottom right, we show a representative STXM contrast image revealing the nanopillar, taken at the Cu L$_3$ resonance energy of 932.7\,eV.}
\label{fig:schematic-measurement}
\end{figure}

Our experimental arrangement, shown in Fig.\,\ref{fig:schematic-measurement}, was similar as that employed in previous studies of spin torque switching \cite{acremann:06}. However, instead of observing directional changes of the large atomic magnetic moments of $\simeq\,2\,\mu_\mathrm{B}$ in a ferromagnetic (FM) Co layer we used the XMCD effect to quantitatively measure the tiny, $\simeq\!10^{-5}\,\mu_\mathrm{B}$/atom, transient signal due to spin currents in interfacial and bulk Cu atoms.

The samples consisted of a multilayer grown by sputtering, where the ferromagnetic polarizer layer [0.3Co/0.9Pd]$_6$/0.3Co/[0.6Ni/0.09Co]$_3$/0.21Co, was designed to have a strong perpendicular magnetic anisotropy and large spin polarization, as previously discussed in \cite{Beaujour:2007}. It is important to note that the Co and Ni layers were deposited at room temperature where they are immiscible so that the resulting Co/Ni multilayer exhibits perpendicular anisotropy and the final interface consists of Co/Cu. The out-of-plane magnetization for the ferromagnetic layer facilitated XMCD imaging in the transmission geometry. The ferromagnetic Co interface layer was followed by a Cu layer of 28 nm. The multilayer stack was fabricated into a nanopillar by using electron beam lithography with a top contact of 5Cr/50Au and a bottom contact of 3Ta/30Ru/3Ta/30Ru/5Ta/2Pd. The pillar had a resistance of $\sim$47$\Omega$. In order to measure the x-ray transmission through the pillar, the Si wafer on the backside was etched leaving the sample supported by the SiN$_{x}$ membrane. The key parts of the sample structure are shown in Fig.~\ref{fig:schematic-measurement}\,(c).

The measurements were performed at the STXM beamline (13-1) at the Stanford Synchrotron Radiation Lightsource (SSRL) shown in Fig.\,\ref{fig:schematic-measurement}\,(b). Circularly polarized x-rays from an undulator were focused to a spot size of 35\,nm by a zone plate lens and the transmitted intensity was measured with an avalanche photo diode. The sample was raster-scanned to generate an image. Current on/off periods applied to the sample were synchronized with the cycle time of the storage ring (781.2 ns) using the counting electronics \cite{Acremann2007}, as illustrated in Fig.\,\ref{fig:schematic-measurement}\,(a). This detection scheme \cite{Bonetti:2015} allowed us to reach a sensitivity of $1\times 10^{-5}$, more than an order of magnitude increase in sensitivity ($6\times 10^{-4}$) over previous attempts to detect spin accumulation with x-rays \cite{Mosendz2009}.

\begin{figure} [!h]
\includegraphics[width=0.8\columnwidth]{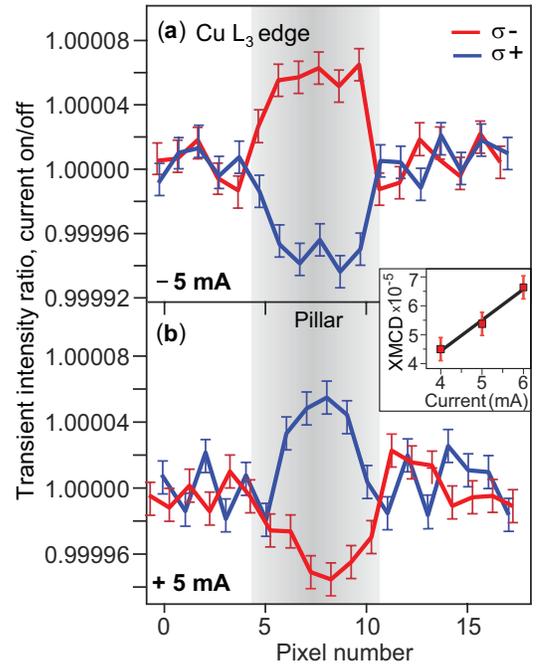}
\caption{ (a) Line scan across the nanopillar. Plotted is the ratio of the transmitted intensities for current on versus off, $I^{\sigma \pm} = I^{\sigma \pm}_\mathrm{on}/I^{\sigma \pm}_\mathrm{off}$, recorded for negative ($\sigma^-$) and positive ($\sigma^+$) helicities of the incident x-rays and -5 mA current. (b) Same as in (a) for opposite current direction (+5 mA). The inset shows the dependence of the XMCD contrast, defined as the intensity ratio $(I^{\sigma +} - I^{\sigma -} $)/($I^{\sigma +} + I^{\sigma -} $).}
\label{fig:images-signal}
\end{figure}

Fig.\,\ref{fig:images-signal}\,(a) shows the averaged line scan across the nanopillar, for the transmitted current on/off intensity ratio, $I^{\sigma \pm} = I^{\sigma \pm}_\mathrm{on}/I^{\sigma \pm}_\mathrm{off}$, recorded with plus ($\sigma +$, blue) and minus ($\sigma -$, red) x-ray helicities. This measurement was taken at -5\,mA, which corresponds to a current density of $10^{11}$A/m$^{2}$. Fig.~\ref{fig:images-signal}\,(b) shows similar line scans for the opposite direction of current flow (+5 mA). The inset shows the current dependence of the XMCD contrast, defined as  $(I^{\sigma +} - I^{\sigma -} $)/($I^{\sigma +} + I^{\sigma -} $). It increases linearly with the current.
The current on/off normalization removes any contrast due to topography and static magnetization. The dependence of the image contrast on x-ray polarization and current flow direction in Figs.~\ref{fig:images-signal}\,(a) and (b) proves that it arises from a magnetic effect. The size of the contrast is similar for both current directions within an error margin of $1\times 10^{-5}$.

\begin{figure} [!h]
\includegraphics[width=0.8\columnwidth]{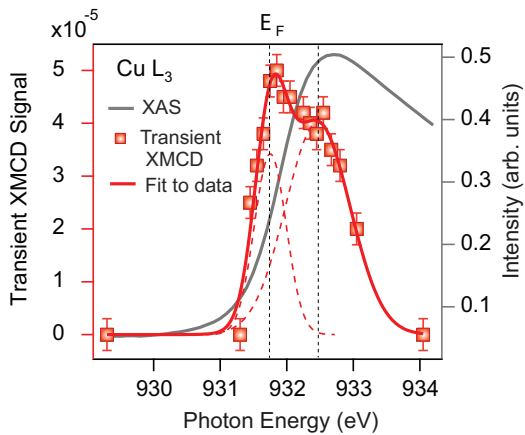}
\caption{Comparison of the L$_3$ x-ray absorption spectrum (XAS) of Cu metal (gray line) with the transient Cu XMCD signal (red squares) of the nanopillar sample, recorded with + 5\,mA current on/off. A fit of the transient spectrum (red line) composed of two Gaussians (dashed red) is shown superimposed.}
\label{fig:energy-dependence}
\end{figure}

Fig.\,\ref{fig:energy-dependence} shows the transient XMCD spectrum (red data points) obtained from integrated image intensities as a function of photon energy across the Cu $L_{3}$ resonance with an applied current of +5\,mA current.
The transient Cu XMCD signal exhibits a two peak structure, clearly revealed by curve fitting with Gaussians (dashed red curves). The lower peak is centered at the inflection point of the Cu metal x-ray absorption spectrum (XAS), shown as a solid gray curve appropriately scaled for reference, and has a full width half maximum (FHWM) of 0.57\,eV. The second peak is located 0.7\,eV higher in energy and has a FWHM of 1.0\,eV.

We assign the lower energy transient XMCD peak in  Fig.\,\ref{fig:energy-dependence}, to Cu atoms in the bulk of the 28\,nm thick film. It is due to spin accumulation induced by a mismatch of the spin dependent resistivities across a Co/Cu interface and exists over a distance that is determined by the spin diffusion length \cite{Silsbee1985}. For our sample, the spin accumulation is approximately constant across the Cu layer whose thickness of 28\,nm  is much smaller than the Cu spin diffusion length of $\simeq \! 350$\,nm \cite{Bass2007,Jedema2001}. Our assignment is supported by the fact, that its position coincides with the inflection point of the Cu metal XAS spectrum which corresponds to the position of the Fermi level $E_\mathrm{F}$ \cite{ebert:96}. The XMCD peak also has the minimum width allowed by the 2p$_{3/2}$  core hole lifetime ($\simeq 0.5$\,eV) \cite{ebert:96, Muller1982}, in good accord with the notion that within the bulk of the Cu film the transient spins flow and accumulate within a narrow energy band around the Fermi level.

The higher energy peak position in Fig.\,\ref{fig:energy-dependence} is close to the peak of the XAS spectrum. Based on earlier work on Co/Cu multilayers (ML) \cite{samant:94, Ruqian1996} and detailed studies of the XMCD and XAS spectra of Co/Cu alloys it is assigned to magnetic Cu interface atoms with a room temperature moment of $\simeq 0.05\,\mu_\mathrm{B}$ \cite{samant:94, Ruqian1996}. For our sample with  a single Co/Cu interface, the static XMCD peak of the interface atoms was too weak to be detected since the Cu signal is dominated by the bulk of the film. However, its \emph{transient} XMCD signal was observable owing to the high sensitivity of our MHz lock-in current on/off technique.

For the same ferromagnetic alignment direction of the Co moments in ML and CoCu alloy reference samples and in our pillar sample, we find that the signs of the static Co and Cu interface XMCD peaks and those of the two transient peaks in Fig.\,\ref{fig:energy-dependence} are the same for a +5mA current direction. From the size of the integrated transient Cu XMCD signal we can estimate the magnetic moment per Cu atom caused by the spin current using the procedure in Ref.\,\cite{samant:94}. Exploiting the fact that the density of states for pure Cu \cite{ebert:96} and a Cu layer sandwiched between Co \cite{nilsson:96} exhibits more $d$ than $s$ states around $E_\mathrm{F}$ and that the XMCD signal is dominated by $2p_{3/2} \rightarrow 3d$ transitions, we derive a moment of $m_\mathrm{Cu} \! \simeq  3 \! \times \! 10^{-5}\,\mu_\mathrm{B}$ per Cu atom for the lower energy peak. If we assign the intensity of the second peak to a single layer of Cu interface atoms, the transient moment per atom is $m_\mathrm{Cu} \! \simeq  4 \! \times \! 10^{-3}\,\mu_\mathrm{B}$.

\begin{figure} [!t]
\includegraphics[width=0.8\columnwidth]{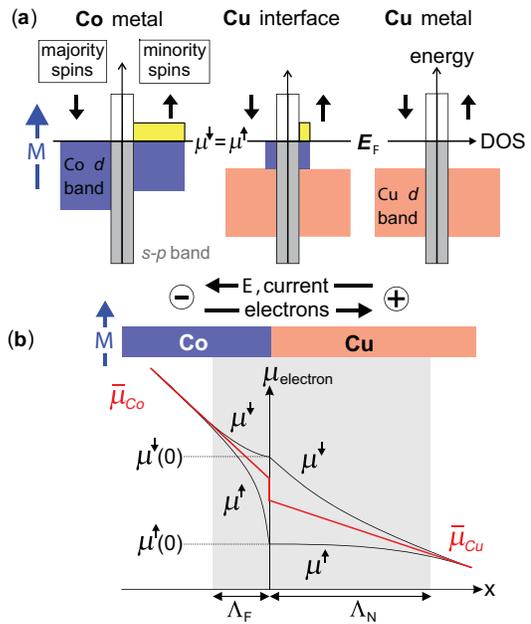}
\caption{
(a) Schematic density of states (DOS) without current flow ($E_\mathrm{F}\!=\!\mu^\uparrow\!=\!\mu^\downarrow$) for itinerant $s$-$p$ and localized $d$ spins for bulk Co and Cu, and Cu interface atoms. The exchange-split $d$ DOS of the Co and magnetic Cu interface atoms exhibits occupied (blue) and unoccupied minority spin states (yellow), which is absent in bulk Cu. The static XMCD effect arises from transitions to the yellow shaded unoccupied $d$ states. (b) Model of the spin dependent electron chemical potentials in the presence of electron spin flow from Co to Cu across an abrupt interface without interface states. The diagram  corresponds to the shown Co magnetization direction and +5mA current as in Fig.\,\ref{fig:schematic-measurement}\,(c). The spin averaged chemical potentials $\bar{\mu}_\mathrm{Co}$ in Co and $\bar{\mu}_\mathrm{Cu}$ in Cu are shown in red. The chemical potentials decay exponentially with distance from their maximum values $\mu^\uparrow(0)$ and $\mu^\downarrow(0)$ at the interface. The origin of the transient XMCD effect is discussed in the text.}
\label{fig:theory}
\end{figure}

The size and sign of the transient Cu moment of the lower energy XMCD peak in  Fig.\,\ref{fig:energy-dependence}, which is assigned to spin accumulation in the bulk of the 28\,nm thick Cu film, can both be explained by Mott's two current model \cite{Mott1936, Fert1968}. In this model the current flows in independent, parallel spin-up and down channels, and spin-flip scattering is forbidden. In each spin channel, the resistivity is determined by scattering of itinerant $s$-$p$ electrons into empty $d$ states localized on the atomic sites. In a strong ferromagnet like Co, the majority $d$ states lie below the Fermi energy and are filled. The resistivity is therefore determined by transitions of minority $s$-$p$ electrons into minority $d$ holes in accordance with the band structure schematically illustrated in Fig.\,\ref{fig:theory}\,(a). In Cu, the $d$ band lies well below the Fermi level  and the lack of localized empty $d$ states in both spin channels leads to a low resistivity.

When electrons flow from Co to Cu, the minority spins experience a lower resistance in Cu since there are less empty $d$ states to scatter into. The Co side of the interface region is therefore preferentially depleted of minority spins. Charge neutrality then requires \emph{accumulation of the majority spins}. Since there must be continuity in each spin channel across the interface, the accumulation of \emph{majority spins} exists on both sides of the interface, as illustrated in Fig.\,\ref{fig:theory}\,(b). For electrons flowing from Cu to Co, the minority spins experience a higher resistance when entering Co. This leads to accumulation of \emph{minority} spins near the interface.

The sign and magnitude of the transient Cu XMCD intensity can be estimated by assuming a change of the spin dependent resistivities and chemical potentials at an interface between two  bulk-like layers \cite{wyder:87}. The magnetic moment due to spin accumulation near a ferromagnet (F) and non-magnet (N) interface is derived in Ref.\,\cite{stohr-siegmann} and for $\ell=$\,F or N is given by,
\begin{eqnarray}
\hspace*{-10pt}m_\ell
\!\!\!&=&\!\!\! - D_\ell(E_\mathrm{F}) \left [\mu^\uparrow (0) -
\mu^\downarrow (0) \right]\,\frac{1}{d}\int_0^d e^{-x/\Lambda_\ell}\, \mathrm{d} x \, \mu_\mathrm{B}
\nonumber \\
\!\!\!&=&\!\!\! D_\ell(E_\mathrm{F}) \frac{2 (2\alpha_\mathrm{F}\! - \! 1)\,\Lambda_\mathrm{N} \, \rho_\mathrm{N}\, e j}{1 \! + \!
4\alpha_\mathrm{F}(1 \!- \!\alpha_\mathrm{F}) \frac{\Lambda_\mathrm{N} \rho_\mathrm{N}}
{\Lambda_\mathrm{F} \rho_\mathrm{F} }}  \, \frac{\Lambda_\ell}{d}  [1\!-\!e^{-d/\Lambda_\ell}] \, \mu_\mathrm{B}
\label{Eq:final-moment-expression}
\end{eqnarray}
We can estimate the transient Cu moment for our layer thickness $d_\mathrm{Cu}\! = \! 28$\,nm. We assume that the spin current is due to the final Co layer from which electrons flow into Cu and use the following values for the spin dependent parameters for bulk Cu and Co \cite{Bass2007}: the densities of states at the Fermi level $D_\mathrm{Co}(E_\mathrm{F})\! = \!0.8$\,atom$^{-1}$eV$^{-1}$ and $D_\mathrm{Cu}(E_\mathrm{F})\! = \!0.2$\,atom$^{-1}$eV$^{-1}$, the spin asymmetry parameter for Co conduction $\alpha_\mathrm{Co}\! = \! 0.8$, the spin diffusion lengths $\Lambda_\mathrm{Co}=38$\,nm and $\Lambda_\mathrm{Cu}\! = \!350$\,nm, and the resistivities $\rho_\mathrm{Co}\! = \! 210\,\Omega$nm  and $\rho_\mathrm{Cu}\! = \! 17\,\Omega$nm. For $j \! = \!+ 1 \!\times \!10^{11}$\,A/m$^2$, we obtain the transient Cu moments to be $m_\mathrm{Cu} \simeq 9.3 \!\times\! 10^{-5}\,\mu_\mathrm{B}$, a factor of three larger than our experimental value of $3 \!\times\! 10^{-5}\,\mu_\mathrm{B}$.

Our model explains the sign reversals with the direction of current flow shown in Fig.\,\ref{fig:images-signal} and accounts for the fact that the moment due to spin accumulation on ``bulk'' Cu atoms revealed by Fig.\,\ref{fig:energy-dependence} has the same sign (direction) as the \emph{static} moments in the previously measured alloys \cite{samant:94}. For the Co magnetization direction in Fig.\,\ref{fig:theory}\,(a), the directions of the static Co and Cu interface moments are parallel, since in both cases there are more minority $d$ holes, identified by yellow shading. The sign of the spin accumulation moment in the bulk of the Cu film follows from Fig.\,\ref{fig:theory}\,(b), which reveals a lowering of the chemical potential for the minority spins resulting in a surplus of minority holes. Hence in all cases the moment direction is the same.

The higher energy peak in Fig.\,\ref{fig:energy-dependence} cannot be explained within the above model which ignores the existence of interface states. Our results reveal a current-induced transient increase of $m_\mathrm{Cu} \! \simeq  4 \! \times \! 10^{-3}\,\mu_\mathrm{B}$ or about 10\% of the 0.05\,$\mu_\mathrm{B}$ static magnetic moment per Cu interface atom. Although the two peaks have about the same size in in Fig.\,\ref{fig:energy-dependence}, the relative abundance of bulk and interface Cu  atoms means that the moment change per Cu atom is about two orders of magnitude larger for the interface ($4 \! \times \! 10^{-3}\,\mu_\mathrm{B}$) than the bulk ($3 \! \times \! 10^{-5}\,\mu_\mathrm{B}$) Cu atoms.

We attribute the large difference in the transient moments to the fact that the interface Cu atoms themselves are magnetic through direct hybridization of their $d$ states with Co \cite{samant:94, Ruqian1996}, yet are more sensitive to thermal and spin current effects than the Co moments. Indeed, the sign of the second peak in Fig.\,\ref{fig:energy-dependence} corresponds to a 10\% \emph{increase} of the Cu interface moments in the current-on relative to the current-off cycle.  Current pulsing leads to an average temperature increase in the pillar of order 100\,K above room temperature. Due to cooling time scale being longer than the duration of current off cycle, the sample remains above room temperature even during the current-off cycle. While the Co magnetization remains stable within the temperature range $<400$\,K, the Cu interface moments are expected to decrease \cite{scherz:2005}. When the current is on, we do not observe a further decrease of the Cu interface moments due to Ohmic heating but rather a spin current induced increase. We attribute it to a spin-torque alignment or stabilization of the Cu interface moments. Finally we mention that conservation of angular momentum during the interfacial spin torque process will lower the spin current polarization in the bulk of the Cu film, in accord with the lower measured than calculated value.

\begin{acknowledgments}
After submission of our paper we became aware of related spin-pumping XMCD studies by J. Li \emph{et al.} http://arxiv.org/abs/1505.03959. Research at SLAC was supported through the SIMES Institute which like the SSRL user facility is funded by the Office of Basic Energy Sciences of the U.S. Department of Energy. Work done at NYU was supported by NSF-DMR-1309202. S. Bonetti acknowledges support from the Knut and Alice Wallenberg Foundation.
\end{acknowledgments}

\end{document}